\documentclass{JHEP3}
\usepackage{graphicx}
\usepackage{amssymb}
\usepackage[mathscr]{eucal}
\usepackage{amsmath}
\usepackage{cite}
\usepackage{afterpage}
\usepackage{slashed}
\usepackage{feynmp}

\newcommand{\gsimm}{\raise.3ex\hbox{$>$\kern-.75em\lower1ex\hbox{$\sim$}}}
\newcommand{\lsimm}{\raise.3ex\hbox{$<$\kern-.75em\lower1ex\hbox{$\sim$}}}

\newcommand{\be}{\begin{equation}}
\newcommand{\ee}{\end{equation}}
\newcommand{\ba}{\begin{eqnarray}}
\newcommand{\ea}{\end{eqnarray}}

\newcommand{\bea}{\begin{eqnarray*}}
\newcommand{\eea}{\end{eqnarray*}}

\title{Explaining the Proton Radius Puzzle with Disformal Scalars}
\author{Philippe Brax \\
  Institut de Physique Th\'eorique, CEA, IPhT, CNRS, URA 2306,
  F-91191Gif/Yvette Cedex, France \\ E-mail:
  \email{philippe.brax@cea.fr}}
 \author{Clare Burrage\\
 School of Physics and Astronomy, University of Nottingham, Nottingham, NG7 2RD, United Kingdom
  \\ E-mail:
  \email{Clare.Burrage@nottingham.ac.uk} }

\date{today}
\abstract{We analyse the consequences of a disformal  interaction between a massless scalar and  matter particles in the context of atomic physics.  We focus on the displacement  of the atomic energy levels that it induces, and
in particular the change in the Lamb shift between the 2s and 2p states. We find that the correction to the Lamb shift depends on the mass of the fermion orbiting around the nucleus, implying a larger effect
for muonic atoms. Taking the cut-off scale describing the effective scalar field theory close to the QCD scale,
we find that the disformal interaction can account for
the observed difference in the proton radius of muonic versus electronic Hydrogen.
Explaining the proton radius puzzle is only possible when  the scalar field is embedded in non-linear theories which alleviate constraints from collider and stellar physics. Short distance properties of the  Galileon where
non-perturbative effects in vacuum are present ensure that unitarity is preserved in high energy particle collisions. In matter,
the chameleon mechanism alleviates the constraints on disformal interactions coming from the burning rates for stellar objects. We show how to combine these two properties in a single model which renders the proposed explanation of the proton radius puzzle viable.
}

\begin{document}

\section{Introduction}

The Lamb shift is one of the most precisely studied  effect in atomic physics. Its relevance has been recently enhanced by the discovery that the Lamb shift behaves differently  when muonic
atoms are considered, compared to their electronic siblings. The Lamb shift can be used to deduce the value of the proton radius and  muonic versus electronic discrepancies imply that
the proton radius is lower by four percent in muonic experiments.   The combined discrepancy between the proton radius as inferred from muonic Hydrogen and that inferred from electronic Hydrogen  now stands at $7\sigma$.  Whilst the muonic results currently only come from one group at PSI no systematic uncertainty has been identified that could explain the size of the discrepancy \cite{Pohl:2010zza}\footnote{There is an ongoing debate on the proton radius discrepancy inferred from e-p scattering experiments \cite{Lorenz:2014vha}.}. This is the proton radius puzzle which has resisted explanation with standard model physics \cite{Pohl:2013yb}.  Could this be an indication of the need for new physics?  Current attempts to explain the proton radius anomaly with new physics have introduced new force carriers with masses in the $1-100 \mbox{ MeV}$ range, which may have non-universal couplings \cite{TuckerSmith:2010ra,Barger:2010aj,Batell:2011qq,Karshenboim:2014tka}, however these are difficult to reconcile with existing constraints on dark forces. New forces deduced from hidden photons or conformally coupled scalars have also been invoked with limited success \cite{Jaeckel:2010xx,Brax:2010gp}. In this work we take a different approach, introducing a nearly massless scalar degree of freedom which interacts with matter species through a universal disformal coupling.

The existence of nearly massless scalar fields is strongly suggested by the acceleration of the Universe, as they could act as dark energy, or arise in theories of massive gravity as
the scalar polarisation of  a low mass  graviton. This seems far removed from the proton radius puzzle, but in this work we will show that these two motivations for considering new physics can be connected.
When conformally coupled to matter, new light scalars are severely constrained by the Cassini probe \cite{Bertotti:2003rm} and tests of the strong equivalence principle such as the Lunar Ranging experiment \cite{Williams:2012nc}. This results in a strong bound on the coupling between the scalar and matter, $\beta$,  that  can, however,  be alleviated  when screening mechanisms are invoked as
non-trivial self-interactions of the field can naturally reduce the strength of the force to observationally undetectable levels in experimental environments. They do this by allowing the properties of the scalar field to vary with the environment. For example; in the chameleon model \cite{KhouryWeltman,Khoury:2003aq} the mass of the scalar field increases in dense environments, in the Galileon model \cite{Nicolis:2008in}, the prefactor of the kinetic term becomes large in the  vicinity of dense sources.  Even for models with screening mechanisms, however, suitably chosen laboratory tests of theories with screening mechanisms can still be constraining; for models such as chameleons, the conformal coupling to matter could be tested in neutron experiments where the energy levels of the neutron in the terrestrial gravitational field are measured \cite{Brax:2011hb}. In this article, we will investigate new tests at the atomic level, due to a disformal coupling between matter and scalars, and we will rely on a screening mechanism to alleviate constraints from higher density environments such as stellar interiors.

Bekenstein has shown \cite{Bekenstein:1992pj} that the most general metric that can be constructed from $g_{\mu\nu}$ and a scalar field that respects causality and the weak equivalence principle is;
\begin{equation}
\tilde{g}_{\mu\nu}=A(\phi,X)g_{\mu\nu} + B(\phi,X) \partial_\mu \phi \partial_\nu \phi\; ,
\label{eq:bekmetric}
\end{equation}
where the first term gives rise to conformal couplings between the scalar field and matter, and the second term is the disformal coupling.  Here  $X=(1/2)g^{\mu\nu}\partial_{\mu}\phi\partial_{\nu}\phi$.  The conformal coupling gives rise to Lagrangian interaction terms of the form
\begin{equation}
\mathcal{L} \supset A(\phi,X) T_{J\mu}^{\mu}\;.
\end{equation}
and the disformal interactions  give rise to Lagrangian interaction terms of the form
\begin{equation}
\mathcal{L} \supset \frac{B(\phi,X)}{2}\partial_\mu \phi\partial_\nu \phi T_J^{\mu\nu}\;.
\label{eq:coupling}
\end{equation}
where $T^{\mu\nu}_J$ is the energy momentum tensor of matter fields in the Jordan frame, defined by the metric $g^J_{\mu\nu}=A(\phi,X)g_{\mu\nu}$.  The conformal coupling gives rise to Yukawa type long range forces between matter fields.
The disformal coupling has no influence on static configurations of matter as no disformal interaction between static non-relativistic objects is generated. This follows from the vanishing of the coupling (\ref{eq:coupling}) when the only non-vanishing component of $T^{\mu\nu}_J $ is $T^{00}_J$ and $\phi$ is static. This can be extended to all the higher order terms involving more than two derivatives of $\phi$ obtained by expanding the matter Lagrangian in perturbations of $B(\phi,X) \partial_\mu \phi \partial_\nu \phi$.  This means that constraints on disformal couplings are much weaker than on their conformal counterparts. The leading disformal  interaction between two static bodies is a quantum effect at one loop \cite{Kugo:1999mf,Kaloper:2003yf,us} which appears at the $1/M^8=B^2(\phi_0,0)$ level where the loop has been calculated in a uniform scalar background $\phi_0$. This gives rise to a potential of the form  $1/M^8 r^7$, which will be analysed and tested here.

 We will find that the proton radius puzzle can be explained using the mass dependent  disformal potential generated at one loop  in $1/M^8r^7$ when $M$ lies close to the QCD scale $M=\Lambda_{QCD}\approx 220 \mbox{ MeV}$.  This is appropriate for a  model which we only require to be  valid at low energies, below the QCD phase transition and after Big Bang Nucleosynthesis (BBN) in cosmology.
This choice of coupling scale would also lead to small anomalous radii for the deuteron\cite{Pohl:2011pxa} and the He nucleus \cite{he}. With new experimental results for deuterons and He nuclei to be soon published, this is a prediction  of our model which will be soon tested \cite{poh}.

In section \ref{sec:scalar}, we recall  salient properties of disformally coupled scalars. In section \ref{sec:atoms}, we apply these results to the proton radius puzzle, determine the required value of the coupling constant $M$ and predict the ensuing deviations for helium and the deuteron. The  disformal coupling, viewed as a higher order operator, would lead to a violation of unitarity at high energy. It would also increase the burning of stars. In section \ref{sec:stars} , we show that the  constraints on disformal couplings coming from the burning of stars can be alleviated in non-linear models for which the chameleon mechanism, whereby the mass of the scalar becomes large in dense environments, prevents the creation of scalars in stellar media.
In section \ref{sec:collider}
we turn to the strong constraints which spring from the absence of any violation of unitarity in particle collision at high energy.  This is alleviated by embedding the disformal coupling in Galileon models which pass these tests by a novel mechanism whereby classical configurations akin to Black Holes in trans-Planckian scattering are formed. We also show how the chameleon mechanism, which is effective in dense environments,  and  the Galileon, which  applies in vacuum such as the ones of atomic and particle physics, are compatible.
 We then conclude in section \ref{sec:conc}.

\section{Light scalar fields}
\label{sec:scalar}

We consider  a scalar field coupled to matter defined by the action
\be
S=\int d^4x \sqrt{-g}\left(\frac{R}{16\pi G_N} -\frac{1}{2} (\partial \phi)^2\right )  + S_m(\psi_i, \tilde g_{\mu\nu})\;,
\label{eq:action}
\ee
where the  metric governing the interactions between the scalar and matter is given by:
\be
\tilde g_{\mu\nu}= A(\phi) g_{\mu\nu} + B(X) \partial_\mu\phi \partial_\nu \phi\;.
\label{eq:tildemetric}
\ee
This is not the most general scalar metric as given by Bekenstein in equation (\ref{eq:bekmetric}), however it describes all the leading order effects of the disformal and conformal couplings.

 The metric $\tilde{g}_{\mu\nu}$ is the  metric with respect to which matter is conserved.  We impose that the conformal coupling function $A(\phi)$ is the only source of (soft) breaking of the shift symmetry $\phi\to\phi+c$, which  forces the coupling $B(X)$ to be independent of $\phi$. We take a conformal coupling to matter of the form,
\be
A(\phi)=1+ \frac{\beta\phi}{m_{\rm Pl}}\;,
\ee
which is the lowest order breaking of the shift symmetry and we assume that the disformal coupling function can be expanded as
\be
B(X)= \frac{1}{M^4}\left(1+ \sum_{n\ge 1} a_n \frac{X^n}{M^{4n}}\right)\;.
\ee
As $M$ will be the lowest energy scale in the disformal sector of our theory we take the cut-off scale, that defines the model as an effective theory at low energy, to lie just above the  scale $M$. We assume a hierarchy between the scales $M$ and $m_{\rm Pl}/\beta$.  As we will find that $M$ is of the order the QCD scale,  and assuming that  $\beta \sim \mathcal{O}(1)$, protecting these scales is similar to the hierarchy problem of the Standard Model.  We have nothing to add to the solutions of this problem except to note that hierarchies between disformal and conformal coupling scales arise naturally  in theories of massive gravity \cite{deRham:2010kj}.
When $\beta \sim \mathcal{O}(1)$  in laboratory interactions the conformal coupling is so weak that it can be safely neglected. The disformal coupling scale $M$  appears as a  one-loop interaction. For  matter sources of masses $m_1$, $m_2$  separated by a distance $r$ the potential interaction mediated by the disformal scalars is \cite{Kaloper:2003yf,us}
\be
V(r)=- \frac{3m_1m_2}{32\pi^3 r^7 M^8}\;,
\ee
when the scalar is canonically normalised.
The coupling scale $M$ is in principle unknown and should be fixed by observations. Here we will focus on theories where this scale is close to  the QCD scale
\be
M\sim\Lambda_{QCD}\;,
\ee
where $\Lambda_{QCD}= 217_{-23}^{+25}$ MeV is the strong interaction scale of quantum chromodynamics (QCD). This choice is compatible with the desire to view our disformal scalars as a low  energy description of some unknown physics which should appear for scales larger than $M$. Below the scale $M$, the physics only involves matter particles which are the electrons, the protons and the neutrons as formed during Big Bang Nucleosynthesis (BBN).
At higher energies, the model must either be completed by some new Ultra Violet (UV) physics, or as we shall see with the example of the Galileon, enter a new phase of the model where perturbative calculations fail and non-perturbative phenomena should be taken into account.

Experimental constraints on disformal couplings have been extensively studied in \cite{us} and these are reproduced in Table \ref{tab:summary}.  We will discuss in later sections how to make our requirement for $M\sim \Lambda_{QCD}$ compatible with all current constraints.
\begin{table}
\begin{tabular}{ |c | c | c |  }
  \hline
  Source of bound & Lower bound on $M$ in GeV & Environment  \\
	  \hline
		Unitarity at the LHC & 30 &Lab. vac. \\
		CMS mono-lepton & 120 &Lab. vac. \\
   CMS mono-photon & 490 &Lab. vac.  \\
  Torsion Balance & $7 \times 10^{-5}$  & Lab. vac. \\
Casimir effect & 0.1 & Lab. vac.  \\
  Hydrogen spectroscopy & 0.2  & Lab. vac.   \\
 Neutron scattering & 0.03  & Lab. vac. \\
Bremsstrahlung & $4 \times 10^{-2}$ & Sun \\
  & 0.18  & Horizontal Branch \\
Compton Scattering & 0.24  & Sun\\
 & 0.81 & Horizontal Branch \\
Primakov & $4\times 10^{-2}$  & Sun \\
   & 0.35 & Horizontal Branch \\
Pion exchange & $\sim 92$  & SN1987a \\
  \hline
\end{tabular}
\caption{Summary of the constraints on the disformal coupling scale $M$ derived in \cite{us}. Lab. vac. means the constraint derives from an experiment conducted in a  laboratory vacuum on Earth.  Horizontal branch means the constraint derives from observations of horizontal branch stars, and similarly for constraints labelled Sun and Supernova SN1987a. }
\label{tab:summary}
\end{table}

In what follows we will restrict ourselves to the leading order effects of the disformal coupling between the scalar field and matter.  Therefore we calculate only to leading order in $1/M^4$, implying  the action can be expanded as
\be
S=\int d^4x \sqrt{-g}\left(\frac{R}{16\pi G_N} -\frac{1}{2} (\partial \phi)^2 + \frac{1}{M^4} \partial_\mu\phi\partial_\nu\phi T_J^{\mu\nu}\right) + S_m(\psi_i, A(\phi)g_{\mu\nu})\;,\label{eq:GravityFieldAction}
\ee
where we have introduced the Jordan frame energy-momentum tensor
\be
 T^{\mu\nu}_J= \frac{2}{\sqrt{- g_J}} \frac{\delta S_m}{\delta  g^J_{\mu\nu}}\;.
\ee
Notice that this last action is written in the Einstein frame and involves the coupling between the Jordan frame energy-momentum tensor $T_J$ and the scalar derivatives.

\section{Microscopic effects}
\label{sec:atoms}
It has been shown that the disformal coupling to matter induces a one loop potential between matter particles \cite{Kaloper:2003yf,us}. This potential is highly sensitive to short distances as it scales as $1/r^7$. Atomic physics experiments are therefore ideal settings to test the influence of the disformal coupling on the properties of matter. As the conformal coupling scale is $\mathcal{O}(m_{\rm Pl})$ it can be safely neglected over atomic distance scales.   In previous work we have shown that the strongest constraints on $M$ from such experiments comes from precision spectroscopy of Hydrogen atoms and constrains the scale $M \gtrsim 200 \mbox{ MeV}$.   In this Section we will determine whether disformal couplings satisfying this bound can explain the proton radius anomaly.  This requires determining disformal corrections to the Lamb shift in Hydrogen from which the proton radius can be inferred.
We will find that  the disformal Lamb shift  is  sensitive to the mass of the particle orbiting around the atomic nucleus, hence inducing different effects in muonic compared to electronic atoms.

\subsection{Lamb shift and proton radius}

The scalar interaction due to the one loop effect of the disformal coupling to matter acts as a perturbation of the Coulombic interaction in Hydrogen-like atoms
\be
V(r)= -\frac{ e^2}{r}  -\frac{3m_1m_2}{M^8} \frac{1}{32\pi^3 r^7}\;.
\ee
In first order perturbation theory, the atomic levels are perturbed by
\be
\delta E=-\frac{3m_fm_N}{32\pi^3 M^8} \left\langle E\left\vert \frac{1}{ r^7}\right\vert E\right\rangle\;,
\ee
 where $\vert E\rangle$ is the unperturbed wave function of a given  energy level.
 Let us focus on Hydrogen-like atoms and consider the 2s and 2p levels, as used to calculate the Lamb shift. In each case the perturbed energy levels are sensitive to the small $r$ parts of the wave function, $r \ll a_0$ where $a_0$ is the Bohr radius and read
 \begin{eqnarray}
 \psi_{2s}(r) & \approx & \frac{1}{2\sqrt{2\pi}} \left(\frac{Z}{a_0}\right)^{3/2};,\nonumber \\
 \psi_{2p}(r) & \approx & \frac{1}{\sqrt{\pi}} \left(\frac{Z}{2a_0}\right)^{5/2} r \cos\theta;,\nonumber
 \end{eqnarray}
 resulting in the perturbation of the 2s and 2p levels given by
  \be
  \delta E_{2s}= -\frac{3}{248\pi^3} \left(\frac{Z}{a_0}\right)^{3} \frac{m_N m_f}{M^8 r_N^4};,
 \ee
 and
 \be
 \delta E_{2p}= -\frac{1}{2^9  \pi^3} \left(\frac{Z}{a_0}\right)^{5} \frac{m_N m_f}{M^8 r_N^2};,
 \ee
 where the interaction has been cut-off at the nuclear radius $r_N$ as below this scale the internal structure of the nucleus becomes relevant.
This leads to a contribution to the Lamb shift $\delta E_{2s-2p}= \delta E_{2p}- \delta E_{2s}$ which is given by:
 \be
 \delta E_{2s-2p}=\frac{3}{248 \pi^3} \left(\frac{Z}{a_0}\right)^{3} \frac{m_N m_f}{M^8 r_N^4}\left[1- \frac{1}{6} \left(\frac{Z}{a_0}\right)^{2} r_N^2\right]\;.
 \ee
The Lamb shift can be used to infer the proton radius $r_N=r_p$ in atoms where the nucleus reduces to a single proton. The phenomenological parametrisation of the Lamb shift in terms of QED and nuclear physics effects and their dependence on the radius $r_p$ is given by \cite{Pohl:2010zza}
\be
\frac{\Delta E_{2s-2p}}{\rm meV}= 210 -5.23 \left(\frac{r_p}{\rm fm}\right)^2 + 0.035 \left(\frac{r_p}{\rm fm}\right)^3\;.
\ee
A new interaction, such as the disformal one, would lead to  a change in the Lamb shift $\delta E_{2s-2p}$  which would be read as a corresponding change in the proton radius $\delta r_p$:
\be
\frac{\delta E_{2s-2p}}{\rm meV}=-10.31 \frac{\delta r_p}{r_p}\;.
\ee
Experimentally, the proton radius deduced from electronic Hydrogen measurements is given by $ r_p= 0.8758 (77) \rm{fm}$, and this  agrees with the charge radius obtained in  electron scattering experiments at low energy \cite{Pohl:2013yb}.
The same measurements of the Lamb shift  can be conducted for muonic atoms, and surprisingly the proton radius appears to be significantly lower;  $r_p=0.84087(39)$ fm, \cite{Pohl:2010zza}, representing a decrease of approximately  four percent.

Reinterpreting the disformal contribution to the Lamb shift as a change in the proton radius for muonic Hydrogen gives
\be
\left(\frac{\delta r_p}{r_p}\right) =  -0.90 \left(\frac{217 \mbox{ MeV}}{M}\right)^8\;,
\ee
where we have taken $r_N$ to be the unperturbed proton radius. The electron contribution is suppressed compared to the muonic contribution by the ratio of electron to muon masses.
Therefore to account for a four percent shift in the proton radius in muonic Hydrogen we must choose:
\begin{equation}
M= 320 \mbox{ MeV}\;,
\label{eq:M}
\end{equation}
which  lies close to the QCD scale. This is compatible with constraints from measurements of Hydrogen spectroscopy which require $M\gtrsim  200 \mbox{ MeV}$.  It is possible to explain the proton radius puzzle because muons orbit closer to the nucleus of an atom than electrons and the disformal force strengthens rapidly with decreases in distance.

One could ask why should the  value of $M$  lie close to the QCD scale?
We can only give a plausibility argument: we want to describe low energy physics in the late universe.  As we are not sensitive to internal nuclear structure it makes sense to cut off the physical description around $\Lambda_{QCD}$.  We also should not be sensitive to physics in the early universe.  The  averaged density of the earth, around $5\ {\rm g cm^{-3}}$, correspond to the densities during BBN.
In this environment where the density is similar to the density of the universe during BBN, it is likely that the cut-off scale $M$ should be close to the cut-off scale of the particle physics model during BBN, i.e. $\Lambda_{QCD}$.

\subsection{The helium radius}
We can extend our study to  He ions carrying one muon compared to those with  one electron. In this case the Lamb shift is related to the Helium radius as \cite{he}
\be
\frac{\Delta E_{2s-2p}}{\rm meV}= 1670.37 -105.322\left(\frac{r_{He}}{\rm fm}\right)^2 + 1.529 \left(\frac{r_{He}}{\rm fm}\right)^3\;.
\ee
The disformal interaction  would induce a change in the Lamb shift,  which is connected to a change in the He radius in the following way:
\be
\frac{\delta E_{2s-2p}}{\rm meV}=-573.5 \frac{\delta r_{He}}{r_{He}}\;.
\ee
Using a value of $M$ determined in Equation (\ref{eq:M}),  $r_{He}=1.681$ fm with $m_{He}=3.728$ GeV and $Z=2$ for the two protons in the He nucleus, we find that the disformal coupling would
induce a change in the Helium radius
\be
\frac{\delta r_{He}}{r_{He}}=0.2\% \;.
\ee
This is smaller than the uncertainty in the Helium radius coming from   $e$ scattering experiments, which is of order $0.3 \%$, and therefore is not currently a detectable effect.

\subsection{The deuterium}

New experiments will also give their results on the deuteron's radius as inferred from the Lamb shift in  muonic deuterium. The deuteron has a  radius of $r_d=2.1402$ fm and a mass $m_d=1.875$ GeV. This leads to a shift in the muonic case of the energy levels
\be
{\delta E_{2s-2p}}=0.023\ {\rm meV}\;.
\ee
This is a prediction of our model which should be compared with future experimental results.

\section{Stellar Burning Constraints}
\label{sec:stars}

\subsection{Constraints from stars}

We have just seen that the proton radius puzzle seems to indicate that the scale $M\sim \Lambda_{QCD}$. This is a low energy scale and one may wonder if the  disformal interaction
may not have an influence on the burning rate of stars, as is the case for axions and axion-like particles. These constraints are summarised in Table \ref{tab:summary}.  For disformally coupled scalar fields the light particles could be emitted by processes such as Compton scattering, bremsstrahlung or Primakov processes in stars of the main sequence and on the horizontal branch of the Hertzsprung-Russell diagram. Two scalars would also be emitted by nuclear processes involving the pion exchange in supernovae. The latter process gives the most severe constraints but suffers from theoretical uncertainties due to the
fact that the pion exchange diagram between two nuclei is a strongly coupled effect treated at tree level, although  higher order effects could alter the result drastically. In addition the maximal emissivity of supernovae $\epsilon_{SN}\lesssim  10^{19} {\rm erg/g\cdot s}$ as
deduced from the SN1987A explosion is only a rough estimate \cite{Raffelt:1996wa}. The constraints from these processes  have been presented in \cite{us} and it was shown  that for the sun we must impose that $M \gtrsim 240$ MeV, for horizontal branch stars
$M\gtrsim 810$ MeV and for supernovae $M\gtrsim 92$ GeV. The solar constraint is always satisfied if we take $M\sim 320$ MeV. The horizontal branch and supernovae constraints need to be analysed carefully as they rule out, at face value, a disformal explanation of the proton radius puzzle.

\subsection{Alleviating the burning constraints with chameleons}

We now embed the disformally coupled scalar field in a chameleon model. These are scalar field theories where a potential term $V(\phi)$ is added to the Lagrangian
\be
S=\int d^4x \sqrt{-g}\left(\frac{R}{16\pi G_N} -\frac{1}{2} (\partial \phi)^2-V(\phi)\right )  + S_m(\psi_i, \tilde g_{\mu\nu})\;.
\label{eq:action1}
\ee
Working to leading order in the disformal coupling, this becomes the effective action
\be
S=\int d^4x \sqrt{-g}\left(\frac{R}{16\pi G_N} -\frac{1}{2} (\partial \phi)^2 -V(\phi)+ \frac{1}{M^4} \partial_\mu\phi\partial_\nu\phi T_J^{\mu\nu}\right) + S_m(\psi_i, A(\phi)g_{\mu\nu})\;.
\ee
One of the salient features of these scalar-tensor models is that, in the presence of non-relativistic matter with a density $\rho$, the dynamics of the scalar field are governed by the effective potential
\be
V_{\rm eff}(\phi)= V(\phi) +(A(\phi)-1)\rho\;,
\ee
which appears in the Klein Gordon equation
\be
\Box \phi -\frac{2}{M^4} D_\mu(\partial_\nu T^{\mu\nu}_J)=\frac{\partial_\phi V_{\rm eff}(\phi)}{\partial\phi}\;.
\ee
When the matter density is constant inside a dense region of the Universe, the field settles at the minimum of the effective potential $\phi(\rho)$ satisfying
\be
\left.\frac{\partial_\phi V_{\rm eff}(\phi)}{\partial\phi}\right\vert_{\phi(\rho)}=0\;,
\ee
when it exists, e.g. for decreasing potentials $V(\phi)$ and increasing coupling functions $A(\phi)$. When the mass of the scalar field, defined as
\be
\left.m^2(\rho)\equiv \frac{\partial_\phi^2 V_{\rm eff}(\phi)}{\partial\phi^2}\right\vert_{\phi(\rho)}\;,
\ee
increases with the matter density $\rho$, the model is a chameleon theory which evades gravitational tests on scalar fifth forces when  the scalar interaction range $\lambda (\rho)=m^{-1}(\rho)$  becomes small enough in dense environments.

Chameleon models
are fully characterised by   their mass  as a function of the density $\rho$ and the coupling $\beta$. There is a one to one correspondence between $m(\rho)$ and $V(\phi)$ \cite{Brax:2011aw} which allows one to define models \be
m(\rho)\sim m_0 \left(\frac{\rho}{\rho_0}\right)^{(n+2)/2}\;,
\ee
where $n>0$ is an index which fully characterises the model. This is, for instance, the mass function of $f(R)$ models in the Einstein frame and the large curvature regime.
The density $\rho_0$ is the matter density in the Universe now. Local tests of gravity require that
\be
m_0 \gtrsim 10^3 H_0\sim 10^{-30}\ {\rm eV}\;,
\ee
which is a very low mass, i.e. the scalar is nearly massless in vacuum.  In stars such as the ones on the Horizontal Branch or supernovae, the scalar field settles at its minimum where $\rho/\rho_0 \sim 10^{33}$ and  $\rho/\rho_0 \sim 10^{43}$ respectively.
Taking $n=1$ for instance, we find that the mass of the scalar field in these environments far exceed the temperature and therefore scalars are not created inside stars simply for kinematical reason\cite{Brax:2007ak}.
Hence disformal scalar fields that possess a  chameleon mechanism  evade the constraints on the disformal coupling coming from the burning rate of stars.

\section{Collider Physics Constraints}
\label{sec:collider}

\subsection{Unitarity constraint}

Strong constraints on the scale $M$ can also be obtained from particle physics. Indeed the disformal coupling is nothing but an irrelevant operator of higher order whose presence jeopardises the UV behaviour of the model.
We can  evaluate when this breakdown occurs and above which scale the effective field description must be altered, by  analysing the unitarity of scattering processes. We focus on  high energy physics experiments and consider that $M\approx 320 \mbox{ MeV}$, as suggested by proton radius measurements.
A typical scattering experiment will involve the creation of two scalars from the annihilation of two fermions $f\bar f \to \phi\phi$. The disformal matrix element for this process becomes
\be
{\cal M}= \frac{{ 2\sqrt{2}}  m_f E^3}{M_{}^4}\;,
\ee
in terms of the energy of the incoming particles in the centre of mass frame.
Perturbative unitarity is preserved when ${\cal M}\le 16 \pi$ implying an energy bound
\be
E\le E_{\rm max}=\left(\frac{8\pi M_{}^4}{ \sqrt {2} m_f}\right)^{1/3}\;.
\ee
Unitarity has been precisely tested in the standard model with LEP where $m_f=m_e$ and we find
\be
E_{\rm max}\sim 7 \ {\rm GeV}\;.
\ee
Hence unitarity would have been  violated at LEP, which reached beam energies of $200 \mbox{ GeV}$, if we extrapolate the disformal coupling between scalars and matter to such high energies.  However if we expect the cut-off to lie just above the scale $M_{}$ then we are not able to naively extrapolate the theory to such high energies.  There are two possible ways to proceed.  The first is that, for energies larger than $E_{\rm max}$, the model must be UV completed in the Wilsonian sense, and the disformal interaction replaced by another interaction between new degrees of freedom, replacing the low energy field $\phi$, and unitarity is restored.  However in this case we can make no statement about whether our model is compatible with collider measurements.  The second,  more predictive, alternative is to use the classicalisation property of Galileon models \cite{Dvali:2010jz}, which we will describe in the following section.

\subsection{Galileons}

We embed the disformally coupled scalar field in a Galileon model \cite{Nicolis:2008in}.  These are scalar field theories which have equations of motion that are at most second order in derivatives, despite the presence of non-trivial derivative self-interactions. In flat space the theory also respects the symmetry $\phi\rightarrow \phi + b_{\mu}x^{\mu} +c$ for constant $b_{\mu}$ and $c$.  We work with this simplest form of the theory, the  cubic Galileon, which has the following Lagrangian
\begin{equation}
\mathcal{L} = -\frac{1}{2}(\partial \phi)^2 -\frac{1}{\Lambda^3}\Box\phi (\partial \phi)^2 +\frac{\beta\phi}{m_{\rm Pl}}T+\frac{1}{M^4} \partial_{\mu}\partial_{\nu}T^{\mu\nu}\;,
\end{equation}
In addition to the coupling scales $m_{\rm Pl}/\beta$ and $M$ the theory is determined by  $\Lambda$, a suppression scale which controls the derivative self interactions that define the Galileon. Around a spherically  symmetric source of mass   $m$ the scalar field profile is
\begin{equation}
\frac{d \phi}{dr} = -\frac{\Lambda^3 r}{4} \left( 1-\sqrt{1+\left(\frac{R_*}{r}\right)^3}\right)\;,
\label{eq:vainphi}
\end{equation}
 and  the non-linearities dominate the evolution of the scalar  within the Vainshtein radius
\begin{equation}
R_* =\frac{1}{\Lambda}\left(\frac{\beta m}{2 \pi m_{\rm Pl}}\right)^{1/3}\;.
\label{eq:vainshtein}
\end{equation}
Within this radius the non-linearities act to suppress the scalar force, $F_{\phi}$, compared to that of Newtonian gravity, $F_N$, so that
\begin{equation}
\frac{F_{\phi}}{F_N} = \beta^2 \left(\frac{r}{R_*}\right)^{3/2}\;.
\end{equation}
Outside the Vainshtein radius, the non-linearities in the kinetic terms become irrelevant and the dominant kinetic term reduces to $-(\partial\phi)^2/2$. Inside the Vainshtein radius,
any perturbations around the background of equation (\ref{eq:vainphi}) inherit a wave function renormalisation such that the kinetic terms of the perturbations read $Z\frac{(\partial\delta\phi)^2}{2}$ where we have
\begin{equation}
|Z| \sim 1 + \frac{ \phi^{\prime}}{r \Lambda^3}\;,
\end{equation}
 and a prime denotes a derivative with respect to radius.  Therefore inside the Vainshtein radius $Z$ can be large.

The Galileon models rely on high mass dimension operators and therefore are sensitive to  quantum corrections at short distance.
 At the quantum level, the Galileon models receive corrections which preserve the Galilean symmetry. Many operators which are not present at tree level appear, and the effective action calculated using the one-particle irreducible diagrams for the Galileon  models can be organised in an infinite series which depends only
on the effective cut-off scale
\be
\Lambda_Z = \sqrt{Z}\Lambda\;,
\ee
and its derivative. This follows from the fact that expanding $\phi=\phi_0 +\delta \phi$ and canonically normalising the field $\delta \phi$, the only dimensionful quantities  controlling self interactions of the field are $\Lambda_Z$ and the
effective mass of $\delta\phi$ which depends only on ${\partial \Lambda _Z}$ and its derivatives. As a result, the quantum corrections are given by
\be
\delta {\cal L}= \Lambda_Z^4 F\left( \frac{\partial^i \Lambda_Z}{\Lambda_Z^{i+1}}\right)\;,
\label{va}
\ee
where the function $F$ depends on the multiple derivatives of $\Lambda_Z$. The overall factor $\Lambda_Z^4$ appears for dimensional reason. For instance, at one loop, the mass term behaves like $ m_\phi\sim \frac{\partial \Lambda_Z}{\Lambda_Z}$ and the Coleman-Weinberg potential reads $\delta V \sim m_\phi^4$ which corresponds to $F(x)= x^4$ where $x= \frac{\partial \Lambda_Z}{\Lambda_Z^2}$.

This result has important consequences.
Firstly, the conformal coupling determined by $\beta$ and the disformal coupling scale $M$ are not renormalised by Galileon fluctuations \cite{deRham:2012az}. Secondly,  the structure of $\delta \mathcal{L}$ implies that the Galileon interactions are not renormalised, the corrections appearing with at least one more derivative at the same order in the $\phi$ expansion. The higher order terms from $\delta \mathcal{L}$ are therefore  small compared to the Galileon terms provided that $\frac{\partial^i \Lambda_Z}{\Lambda_Z^{i+1}}\lesssim  1$. Inside the Vainshtein radius the dependence of  $\Lambda_Z$ on $r$ is a power law and therefore this condition is met provided
\be
r\Lambda_Z\gtrsim 1\;.
\ee
For the cubic Galileon, we have inside an object and up to its surface
\be
\Lambda_Z \sim \left(\frac{\beta \bar\rho \Lambda}{m_{\rm Pl}}\right)^{1/4}\;,
\ee
for an object of averaged density $\bar \rho$. This scale is independent of $r$ and guarantees that one can trust the Vainshtein solution as long as
\be
r\gtrsim r_E\sim \left(\frac{m_{\rm Pl}}{\beta \bar\rho \Lambda}\right)^{1/4}\;,
\ee
We have used a free scalar field  model down to the nuclear scale in the disformal calculation of the Lamb shift, so we must impose that $r_N\Lambda_Z\gtrsim 1$. Taking
$\bar \rho \sim r_N^{-4}$ for nuclear matter, we find that $r_E\gg r_N$ for $\Lambda \ll m_{\rm Pl}$ and $\beta={\cal O}(1)$. This would prevent us from trusting our loop calculation
of the Lamb shift. The only possibility is to impose that the Vainshtein radius of nucleons is smaller than their size, implying that the Galileon theory is weakly coupled down to nuclear scales. This requires that
\be
\Lambda \gtrsim \frac{1}{r_N} \left(\frac{\beta m_N}{m_{\rm pl}}\right)^{1/3}\sim 0.2\ {\rm keV}\;.
\ee

\subsection{Non-perturbative effects}

In some models, the perturbative assumption that  single particle states  are created from  annihilation processes is no longer valid in high energy collisions. This happens when non-perturbative effects occur at high energy due to the non-linearities
of the theory. In particular, classical lumps can be created in some scalar models that would be akin to the creation of black holes  at Planck scale energies.
This phenomenon alleviates the perturbative unitarity bound as the growth of the scattering cross section with the energy is modified. Let us illustrate this with the Galileon models which offer the possibility of  annulling the constraint  from perturbative unitarity.

The Vainshtein radius plays an important role in particle scattering experiments.
Take, for instance, the case of a two fermion annihilation process previously discussed. On shell, the trace of the energy momentum tensor of massive fermions is
\be
T=-m_F \bar\psi \psi\;,
\ee
corresponding to a non-vanishing  energy density which is highly concentrated and can be modelled  as
\be
T^{00}=\sqrt s \delta^{(3)}\;,
\ee
in the centre of mass frame, with $s=-p^2$ where $p$ is the total incoming 4-momentum. This is a peaked energy density $\rho$ of order $\sqrt s$ over a region of size $R\sim \frac{1}{\sqrt s}$ during a time interval of order $\Delta t \sim \frac{1}{\sqrt s}$.
During the time interval of the collision when the energy density $\rho$ is non-vanishing, the scalar field can be modelled as a solution of the time independent Klein-Gordon equation sourced by $T^{00}$.  In this frame the tree level  contribution from the disformal term vanishes as the source is static and the scalar field profile is determined by the same equation as the one used to analyse the Vainshtein screening in Galileon models.
The scalar profile becomes a scalar lump with its energy concentrated inside the Vainshtein radius.
The non-linearities dominate inside the Vainshtein radius and are important for scattering experiments provided the source term lies inside its own  Vainshtein radius
$R\lesssim R_\star$. This is the analogue of the criterion for the formations of black holes, i.e. the requirement that the size of the interaction region must be
within its Schwarzschild radius.

Being dominated by the creation of a classical lump, the scattering process has  a total cross section which is of the order of the Vainshtein radius squared
\be
\sigma_T\sim R_\star^2\;.
\ee
 After the time interval $\Delta t$, the energy density drops to zero and the scalar lump cannot be maintained any more. This triggers the classical decay of the  scalar configuration, and the  solution becomes time dependent.  We do not study the details of this decay here, however one expects the energy of the initial lump to spread out in space and decay classically. During this process, one also expect that quantum phenomena take place with the emission of on shell particle states.

For the Galileons, a source of mass $\sqrt s$ and size $R=1/\sqrt s$ has a Vainshtein radius
\be
R_\star (s)= \frac{1}{\Lambda}\left(\frac{\beta \sqrt s}{2 \pi m_{\rm Pl}}\right)^{1/3}\;,
\ee
and the Vainshtein criterion $R\lesssim R_\star(s)$ is satisfied provided
\be
s\gtrsim s_\star\equiv \left(\frac{2\pi \Lambda^3 m_{\rm Pl}}{\beta}\right)^{1/2}\;,
\ee
where $\beta={\cal O}(1)$.
The high energy behaviour above $s_\star$ determined here must resolve the breaking of perturbative unitarity implying that $s_\star\le E_{\rm max}^2\sim (7\mbox{ GeV})^2$ and therefore
\be
\Lambda^3 \le \frac{\beta E_{\rm max}^4}{2\pi m_{\rm Pl}}\;,
\ee
which corresponds to $\Lambda \le 6$ keV.  Unitarity must also be respected in the scalar sector of the theory where the $\phi+\phi\to \phi+\phi$ process leads to a scattering amplitude at tree level ${\cal M}\sim \frac{s^3}{\Lambda^6}$ in a terrestrial environment. At higher energy, one must resum all the ladder diagrams involving the three-point vertex of Galileon models with an amplitude of the form ${\cal M}\sim \frac{s^3}{\Lambda^6}(1+\alpha \frac{s^3}{\Lambda^6})^{-1}$ where $\alpha$ is a numerical factor. This resummation renders the amplitude bounded at large $s$ guaranteeing that unitarity is not violated.

 Whilst this classicalisation process alleviates the unitarity bounds on the disformal coupling, the decay of the scalar lumps must still be compatible with collider observations.  At energies
higher than $\sqrt s_\star$, fermion annihilation including those at  LEP  and the LHC would not create two particle scalar states but  classical Galileon configurations.
These Galileons would then decay  classically and quantum mechanically into a few multi-particle states coming from the
coupling of the scalar field to matter fields.   A full calculation of this process is beyond the scope of this work, however as the coupling between the scalar and matter particles is suppressed by $\beta /m_{\rm Pl}$ for the conformal coupling and by loop suppression factors for the disformal coupling we expect observable signatures of this process to be very difficult to detect.

\subsection{Combining the chameleon and Galileon effects}

We have seen that the unitarity bound from particle physics can be alleviated provided $\Lambda \lesssim 6 $ keV. We have also found that the calculation of the disformal Lamb shift can be trusted provided that $\Lambda \gtrsim 0.2 $ keV. This gives a narrow band of values for $\Lambda\sim 1$ keV. For these values of $\Lambda$, the Vainshtein radius of stars of the main sequence and on the horizontal branch are extremely small compared to their sizes. For supernovae, the Vainshtein radius is of the order of the radius of the core. We have used the chameleon mechanism to tackle the burning rate problem for stars. How can we make the chameleon and Galileon compatible?
We consider the full model  defined by the action
\begin{eqnarray}
S&=&\int d^4x \sqrt{-g}\left(\frac{R}{16\pi G_N} -\frac{1}{2} (\partial \phi)^2 -\frac{1}{\Lambda^3}\Box\phi (\partial \phi)^2-V(\phi)+ \frac{1}{M^4} \partial_\mu\phi\partial_\nu\phi T_J^{\mu\nu}\right) \nonumber\\
& &\; + S_m(\psi_i, A(\phi)g_{\mu\nu})\;,
\end{eqnarray}
where $V(\phi)$ and $A(\phi)$ depend on the chameleon model. Inside a dense body of almost uniform density, the field settles at the minimum of the effective potential $\phi (\rho)$. This is even true inside the would-be Vainshtein radius $R_\star$ of the object as the source term in the Klein-Gordon for the spatial variation of the field $\frac{\partial{V_{\rm eff}}}{\partial\phi}\vert_{\phi (\rho)}$ vanishes altogether.
This implies that the model behaves like a chameleon model inside matter, and the burning rate bounds are evaded provided the mass of the scalar field in dense media is large enough.  On the contrary, in the sparse environments of atomic or particle physics experiments where the ambient density is low, the scalar behaves like a nearly massless field with a background value $\phi_0$ which minimises the effective potential. During the course of a high energy collision, the extreme densities reached in the centre of mass frame during a time $\Delta t\sim 1/\sqrt{s}$ far exceed the background density and
act as a peaked source localised in space. Expanding $\phi=\phi_0 +\delta \phi$, the action reduces to the one of a cubic Galileon model for a nearly massless field sourced by the matter density coming from the particle collision. As a result, the Galileon lumps can be created at high enough energy and unitarity is restored.

In conclusion, we have introduced a class of models where the chameleon mechanism dominates in dense environments where the scalar field is extremely massive and cannot be produced by the particle reactions involving the disformal coupling. This prevents the dramatic increase of the burning rate of star that a disformal model with $M\sim \Lambda_{QCD}$ would entail. On the other hand, in (near) vacuum situations where particle experiments take place, the
Galileon interactions become relevant at short distance when the Galileon coupling is approximately  $\Lambda \sim 1 $ keV.

\section{Conclusion}
\label{sec:conc}
We have considered the effects of a disformal coupling between a massless scalar field and matter in the context of atomic physics. We have shown that the proton radius puzzle, i.e. a difference of four percent between the Lamb shifts of electronic and muonic atoms, can be explained by such a disformal coupling when the cut-off scale of the model for experiments  carried out in the terrestrial environment is close to  the QCD scale. This allows us to predict that the disformal effect on the  He radius should be below  the percent level. These results are only valid when the scalar model is embedded in non-linear models with the chameleon screening mechanism in dense environments. This helps alleviating the constraints coming from stellar burning rates as scalars are too heavy to be created in such environments. At higher energy and in near vacuum, the Vainshtein mechanisms of cubic Galileon models would prevent the violation of unitarity by disformal interactions. These models would be characterised by the production and decay of classical lumps. The determination of signatures for these events would certainly lead to promising tests of the models presented here at the LHC. This is left for future work.

\section{Acknowledgements}

We are extremely grateful to Randolf Pohl for discussion and private communications. We would like to thank Herbert Dreiner for useful communications.  C.B. is supported by a Royal Society University Research Fellowship. P.B.
acknowledges partial support from the European Union FP7 ITN
INVISIBLES (Marie Curie Actions, PITN- GA-2011- 289442) and from the Agence Nationale de la Recherche under contract ANR 2010
BLANC 0413 01.

\end{document}